\documentclass[11pt]{article}
\usepackage{graphicx}
\usepackage{amssymb}

\newcommand{\be}{\begin{equation}}
\newcommand{\ee}{\end{equation}}

\begin{document}

\title{\bf Multifractal analysis of complex networks}

\author{Dan-Ling Wang$^{1,2}$, Zu-Guo Yu$^{1,3}$\thanks{
  Corresponding author, email: yuzg@hotmail.com} and Vo Anh$^1$\\
{\small $^{1}$School of Mathematical Sciences, Queensland University of Technology,}\\
{\small GPO Box 2434, Brisbane, Q4001, Australia.}\\
{\small $^2$School of Mathematics \& Physics, University of
Science \& Technology Beijing, Beijing 10083, China. }\\
{\small$^3$ Hunan Key Laboratory for Computation and Simulation in
Science and Engineering, }\\
{\small Xiangtan University, Xiangtan,  Hunan 411105, China.}
 }
\date{}
\maketitle

\begin{abstract}
Complex networks have recently attracted much attention in diverse
areas of science and technology. Many networks such as the WWW and
biological networks are known to display spatial heterogeneity
which can be characterized by their fractal dimensions.
Multifractal analysis is a useful way to systematically describe
the spatial heterogeneity of both theoretical and experimental
fractal patterns. In this paper, we introduce a new box covering
algorithm for multifractal analysis of complex networks. This
algorithm is used to calculate the generalized fractal dimensions
$D_{q}$ of some theoretical networks, namely scale-free networks,
small world networks and random networks, and one kind of real
networks, namely protein-protein interaction networks of different
species. Our numerical results indicate the existence of
multifractality in scale-free networks and protein-protein
interaction networks, while the multifractal behavior is not
clear-cut for small world networks and random networks. The
possible variation of $D_{q}$ due to changes in the parameters of
the theoretical network models is also discussed.
\end{abstract}


\section{Introduction }

 Complex networks have been studied extensively due to
their relevance to many real-world systems such as the world-wide
web, the internet, energy landscapes, and biological and social
systems.$^{\cite{Song2005}}$

It has been shown that many real complex networks share distinct
characteristics that differ in many ways from random and regular
networks.$^{\cite{LeeJung2006, GuoCai2009}}$ Three fundamental
properties of real complex networks have attracted much attention
recently: the small-world property,$^{\cite{ER1960, Milgram1967}}$
the scale-free property,$^{[6-8]}$ and the
self-similarity.$^{\cite{Song2005}}$ The small-world property
means that the average shortest path length between vertices in
the network is short, usually scaling logarithmically with the
size $N$ of the network.$^{\cite{GuoCai2009}}$ A famous example is
the so-called \emph{six degrees of separation} in social
networks.$^{\cite{Milgram1967}}$ A large number of real networks
are referred to as \emph{scale-free} because the probability
distribution $P(k)$ of the number of links per node (also known as
the degree distribution) satisfies a power law $P(k)\sim
k^{-\gamma }$ with the degree exponent $\gamma $ varying in the
range $2<\gamma <3$.$^{\cite{AJB1999}}$ In view of their
small-world property, it was believed that complex networks are
not self-similar under a length-scale transformation. After
analyzing a variety of real complex networks, Song \textit{et
al.}$^{\cite{Song2005}}$ found that they consist of self-repeating
patterns on all length scales, i.e., they have \emph{self-similar
structures}. In order to unfold the self-similar property of
complex networks, Song \textit{et al.}$^{\cite{Song2005}}$
calculated their fractal dimension, a known useful characteristic
of complex fractal sets,$^{[9-11]}$ and found that the
box-counting method is a proper tool for further investigations of
network properties. Because a concept of metric on graphs is not
as straightforward as the Euclidean metric on Euclidean spaces,
the computation of the fractal dimension of networks via a
box-counting approach is much more complicated than the
traditional box-counting algorithm for fractal sets in Euclidean
spaces. Actually Eguiluz \textit{et al.}$^{\cite{Eguiluz2003}}$
introduced appropriate definitions of dimensions in order to
characterize the fractal properties of complex networks in 2003.
Song \textit{et al.}$^{\cite{Song2007}}$ developed a more involved
algorithm to calculate the fractal dimension of complex networks.
Then Kim \textit{et al.}$^{\cite{Kim2007}}$ proposed an algorithm
by considering the skeleton of networks. Zhou \textit{et
al.}$^{\cite{ZhouJiang2007}}$ proposed an alternative algorithm,
based on the edge-covering box counting, to explore the
self-similarity of complex cellular networks. Later on, a
ball-covering approach$^{\cite{GaoHu2008}}$ and an approach
defined by the scaling property of the volume$^{\cite{GuoCai2009,
Shanker2007}}$ were proposed for calculating the fractal dimension
of complex networks. Recently fractality and percolation
transition,$^{\cite{Rozenfeld2009}}$ fractal
transition$^{\cite{Rozenfeld2010}}$ in complex networks,
properties of scale-free Koch network$^{\cite{LiuKong2010}}$ were
studied.

The tools of fractal analysis provide a global description of the
heterogeneity of an object, such as its fractal dimension. This
approach is not adequate when the object may exhibit a
multifractal behavior. Multifractal analysis is a useful way to
systematically characterize the spatial heterogeneity of both
theoretical and experimental fractal
patterns.$^{\cite{Grassberger1983, Halsy1986}}$ It was initially
proposed to treat turbulence data, and has recently been applied
successfully in many different fields including time series
analysis,$^{\cite{can00}}$ financial modelling,$^{\cite{Anh2000}}$
biological systems$^{[25-35]}$ and geophysical
systems.$^{[36-42]}$ For complex networks, as mentioned above,
through the recent works,$^{[1,3,14-16]}$  it was already a big
step to go from the computation of the fractal dimension of a
geometrical object to that of a network via the box-counting
approach of fractal analysis. Lee and Jung$^{\cite{LeeJung2006}}$
found that the probability distribution of the clustering
coefficient of complex network is best characterized by the
multifractal. Polla \textit{et al.}$^{\cite{Polla2010}}$
introduced a multifractal network generator. In this paper, we
introduce a new box-covering algorithm to compute the generalised
fractal dimensions of a network. This is a step to move from
fractal analysis to multifractal analysis of complex networks.

We first adapt the random sequential box covering
algorithm$^{\cite{Kim2007}}$ to calculate the fractal dimension of
the human protein-protein interaction network as well as that of
its skeleton. We next introduce a box covering algorithm for
multifractal analysis of networks in Section 2. In Section 3, this
algorithm is then used to calculate  the generalized fractal dimensions $%
D_{q}$ of generated examples of three classes of theoretical
networks, namely scale-free networks, small-world networks and
random networks, and one kind of real networks, namely
protein-protein interaction networks of different species. The
methods to generate the theoretical networks are described. The
multifractal behaviour of these networks based on the computed
generalised fractal dimensions $D_{q}$ is then discussed. The
possible variation of $D_{q}$ due to changes in the parameters of
the theoretical network models is also investigated. Some
conclusions are then drawn in Section 4.

\section{Methods}

 In this section, we first introduce the box covering
methods for calculating the fractal dimension of complex networks
and the traditional fixed-size box counting algorithms used for
multifractal analysis. We then present our new approach for
multifractal analysis of complex networks in detail.

\subsection{The box covering methods for calculation of fractal dimension}

 Box covering is a basic tool to estimate the fractal
dimension of conventional fractal objects embedded in the
Euclidean space. The Euclidean metric is not relevant for complex
networks. A more natural metric is the shortest path length
between two nodes, which is defined as the number of edges in a
shortest path connecting them. Shortest paths play an important
role in the transport and communication within a network. It is
useful to represent all the shortest path lengths of a network as
a matrix $D$ in which the entry $d_{ij}$ is the length of the
shortest path from node $i$ to node $j$. The maximum value in the
matrix $D$ is called the network diameter, which is the longest
path between any two nodes in the network. Song \textit{et
al.}$^{\cite{Song2005}}$ studied the fractality and
self-similarity of complex networks by using box covering
techniques. They proposed several possible box covering
algorithms$^{\cite{Song2005}}$ and applied them to a number of
models and real-world networks. Kim \textit{et
al.}$^{\cite{Kim2007}}$ introduced another method called the
random sequential box covering method, which can be described as
follows:

For a given network, let $N_{B}$ be the number of boxes of radius
$r_{B}$ which are needed to cover the entire network. The fractal
dimension $d_{B}$ is then given by
$$
N_{B}\sim {r_{B}}^{-d_{B}}.
$$
By measuring the distribution of $N_{B}$ for different box sizes,
the fractal dimension $d_{B}$ can be obtained by power law fitting
of the distribution. This algorithm has the following
steps:$^{\cite{Kim2007}}$

\begin{enumerate}
\item[(i)] Select a node randomly at each step; this node serves
as a seed which will be the center of a box.

\item[(ii)] Search the network by distance $r_{B}$ from the seed
and cover all nodes which are found but have not been covered yet.
Assign the newly covered nodes to the new box. If no newly covered
nodes have been found, then this box is discarded.

\item[(iii)] Repeat (i) and (ii) until all nodes in the network
have been assigned to their respective boxes.
\end{enumerate}

To obtain the \emph{skeleton} of a complex network, we firstly
need to calculate the edge betweenness of all the edges in this
network. The betweenness $b_{i},$ also referred to as
load,$^{\cite{Kim2007}}$ is defined as
$$
b_{i}=\sum_{j,k\in N,j\neq k}\frac{n_{jk}(i)}{n_{jk}},
$$
where $N$ is the number of nodes, $n_{jk}$ is the number of
shortest paths connecting nodes $j$ and $k$, while $n_{jk}{(i)}$
is the number of shortest paths connecting nodes $j$ and $k$ and
passing through edge $i$. Similar to a minimum spanning tree, a
skeleton is constructed so that edges which have the highest
betweenness and do not form loops are selected.$^{\cite{Kim2007}}$
The remaining edges in the original network are referred to as
shortcuts that contribute to loop formation. In other words, the
distance between any two nodes in the original network may
increase in the skeleton. For example, in the human
protein-protein interaction network, the largest distance between
any two nodes in the original network is 21 while the largest
distance between any two nodes in its skeleton is 27.

As an example, we used the above algorithm to estimate the fractal
dimension of the human protein-protein interaction network as well
as that of its skeleton. The result is shown in Fig. 1. When we
applied the box covering
algorithm on the skeleton, more boxes were needed for each fixed box radius $%
r_{B}$. The increasing rate of the number $N_{B}$ of boxes varies
when the size $r_{B}$ of the box increases. More specifically,
when $r_{B}$ is smaller, the number of boxes needed is not much
different for both the original network and its skeleton; but when
$r_{B}$ is larger, many more boxes are needed to cover the
skeleton than the original network.

\subsection{Algorithms for multifractal analysis of networks}

 Real-world fractals may not be homogeneous; there
is rarely an identical motif repeated on all scales. Two objects
might have the same fractal dimension and yet look completely
different. Real-world fractals possess rich scaling and
self-similarity properties that can change from point to point,
thus can have different dimensions at different scales. The
present paper investigates these properties on complex networks,
we aim to develop an approach for multifractal analysis of complex
networks.

The most common algorithm of multifractal analysis is the
fixed-size box-counting algorithm. $^{\cite{Halsy1986,
YuAnhLau2001b, YuAnhLau2003}}$ For a given probability measure
$0\leq \mu \leq 1$ with support set $E$ in a metric space, we
consider the partition sum
\begin{equation}
Z_{\epsilon }(q)=\Sigma _{\mu (B)\neq 0}[\mu (B)]^{q},\, \label{1}
\end{equation}%
where $q$ is a real number and the sum runs over all different
non-overlapping boxes $B$ of a given size $\epsilon $ in a
covering of the support $E$. It follows that $Z_{\epsilon
}(q)\geqslant 0$ and $Z_{\epsilon }(0)=1.$ The mass exponent
function $\tau (q)$ of the measure $\mu $ is defined by
\begin{equation}
\tau (q)=\lim_{\epsilon \rightarrow 0}\frac{\ln Z_{\epsilon
}(q)}{\ln \epsilon }.  \label{2}
\end{equation}%

The generalized fractal dimensions of the measure $\mu $ are
defined as
\begin{equation}
D_{q}=\frac{\tau (q)}{q-1},~q\neq 1,  \label{3}
\end{equation}%
and
\begin{equation}
D_{1}=\lim_{\epsilon \rightarrow 0}\frac{Z_{1,\epsilon }}{\ln
\epsilon }, \label{4}
\end{equation}%
for $q=1$, where $Z_{(1,\epsilon )}=\Sigma _{\mu (B)\neq 0}\mu
(B)\ln \mu (B) $.

For every box size $\epsilon $, the number $\alpha =\frac{\log \mu
\left( \epsilon \right) }{\log \epsilon },$ also referred to as
the H\"{o}lder exponent, is the singularity strength of the box.
This exponent may be
interpreted as a crowding index of a measure of concentration: the greater $%
\alpha $ is, the smaller is the concentration of the measure, and
vice versa. For every box size $\epsilon $, the numbers of cells
$N_{\alpha }(\epsilon )$ in which the H\"{o}lder exponent $\alpha
$ has a value within the range $[\alpha ,\alpha +d\alpha ]$ behave
like
$$
N_{\alpha }(\epsilon )\sim \epsilon ^{-f\left( \alpha \right) }.
$$
The function $f\left( \alpha \right) $ signifies the Hausdorff
dimension of the subset which has singularity $\alpha $; that is,
$f(\alpha )$ characterizes the abundance of cells with H\"{o}lder
exponent $\alpha $ and is called the \emph{singularity spectrum}
of the measure. The measure $\mu $
is said to be a \emph{multifractal measure} if its singularity spectrum $%
f\left( \alpha \right) \neq 0$ for a range of values of $\alpha $.
The singularity spectrum $f\left( \alpha \right) $ and the mass
exponent function $\tau (q)$ are connected via the Legendre
transform:$^{\cite{Mandelbrot1983}}$
\begin{equation}
\alpha \left( q\right) =\frac{d\tau \left( q\right) }{dq}
\end{equation}%
and%
$$
f(\alpha \left( q\right) )=q\alpha \left( q\right) -\tau \left(
q\right) ,~q\in \mathbb{R}.
$$
Considering the relationship between the mass exponent function
$\tau (q)$
and the generalized dimension function $D_{q}$, the singularity spectrum $%
f(\alpha )$ contains exactly the same information as $\tau (q)$
and $D_{q}$.

The generalized fractal dimensions are estimated through a linear
regression of $[\ln Z_{\epsilon }(q)]/(q-1)$ against $\ln \epsilon
$ for $q\neq 1$, and similarly through a linear regression of
$Z_{1,\epsilon }$ against $\ln \epsilon $ for $q=1$. The value
$D_{1}$ is called the information dimension and $D_{2}$ the
correlation dimension.

For a network, the measure $\mu $ of each box is defined as the
ratio of the number of nodes covered by the box and the total
number of nodes in the network. The fixed-size box-counting
algorithm of Ref. \cite{Kim2007} described above could not be used
to analyze the multifractal behavior of networks directly. Because
the method contains a random process of selecting the position of
the center of each box, this will affect the number of boxes with
a fixed size. Especially, if a node with large degree (a hub) is
randomly chosen, a lot more nodes could be covered, and it is an
efficient way when we produce box covering. However, if a node
with small degree is randomly chosen first, few nodes could be
covered. As a result, the partition sum defined by Eq. (1) will
change each time we proceed with box counting. To avoid this
effect, we propose to take the average of the partition sums over
a large number of times and accordingly modify the original
fixed-size box-counting algorithm into a new method. To our
knowledge, this improvement is the first introduced in this
approach to analyze the multifractal behavior of complex networks.

\textbf{\ }We need to calculate the shortest-path distance matrix
for each network and these matrices are the input data for fractal
and multifractal analyses. We describe the procedure as follows:

\begin{enumerate}
\item[(i)] Transform the pairs of edges and nodes in a network
into a matrix $A_{N\times N}$, where $N$ is the number of nodes of
the network. The matrix $A_{N\times N}$ is a symmetric matrix
where the elements $a_{ij}=0$ or $1$
with $a_{ij}=1$ when there is an edge between node $i$ and node $j$, while $%
a_{ij}=0$ when there is no edge between them. We define that each
node has no edge with itself and accordingly $a_{ii}=0$.

\textbf{Remark}: $A_{N\times N}$ could be the input data for
calculating the degree distribution and characteristic path length
to determine whether the network possesses the properties of
scale-free degree distribution and small-world effect.

\item[(ii)] Compute the shortest path length among all the linked
pairs and save these pairs into another matrix $B_{N\times N}$ .

\textbf{Remark}: In graph theory, calculation of the shortest path
is a significant problem and there are many algorithms for solving
this problem. Here, in our approach, we use Dijkstra's
algorithm$^{\cite{Dijkstra1959}}$ of the Matlab toolbox.
\end{enumerate}

After the above steps we could use the matrix $B_{N\times N}$ as
input data for multifractal analysis based on our \emph{modified
fixed-size box counting algorithm} as follows:

\begin{enumerate}
\item[(i)] Initially, all the nodes in the network are marked as
uncovered and no node has been chosen as a seed or center of a
box.

\item[(ii)] According to the number of nodes in the network, set $%
t=1,2,...,T $ appropriately. Group the nodes into $T$ different
ordered random sequences. More specifically, in each sequence,
nodes which will be chosen as seed or center of a box are randomly
arrayed.\newline \textbf{Remark}: $T$ is the number of random
sequences and is also the value over which we take the average of
the partition sum $\overline{Z_{r}(q)}$. Here in our study, we set
$T=200$ for all the networks in order to compare them.

\item[(iii)] Set the size of the box in the range $r\in \lbrack
1,d]$, where $d$ is the diameter of the network.\newline
\textbf{Remark}: When $r=1$, the nodes covered within the same box
must be connected to each other directly. When $r=d$, the entire
network could be covered in only one box no matter which node was
chosen as the center of the box.

\item[(iv)] For each center of a box, search all the neighbors
within distance $r$ and cover all nodes which are found but have
not been covered yet.

\item[(v)] If no newly covered nodes have been found, then this
box is discarded.

\item[(vi)] For the nonempty boxes $B$, we define their measure as
$\mu (B)=N_{B}/N,\,$ where $N_{B}$ is the number of nodes covered
by the box $B$, and $N$ is the number of nodes of the entire
network.

\item[(vii)] Repeat (iv) until all nodes are assigned to their
respective boxes.

\item[(viii)] When the process of box counting is finished, we
calculate the partition sum as $Z_{r}(q)=\Sigma _{\mu (B)\neq
0}[\mu (B)]^{q}\,$ for each value of $r$.

\item[(ix)] Repeat (iii) and (iv) for all the random sequences,
and take the average of the partition sums
$\overline{Z_{r}(q)}=(\sum^{t}Z_{r}(q))/T,$ and then use
$\overline{Z_{r}(q)}$ for linear regression.
\end{enumerate}

Linear regression is an essential step to get the appropriate
range of $r\in
\lbrack r_{min},r_{max}]$ and to get the generalized fractal dimensions $%
D_{q}$. In our approach, we run the linear regression of $[\ln \overline{%
Z_{r}(q)}]/(q-1)$ against $\ln (r/d)$ for $q\neq 1$, and similarly
the linear regression of $\overline{Z_{1,r}}$ against $\ln (r/d)$
for $q=1$, where $\overline{Z_{1,r}}=\Sigma _{\mu (B)\neq 0}\mu
(B)\ln \mu (B)$ and $d$ is the diameter of the network. An example
of linear regression for the Arabidopsis thaliana PPI network is
shown in Fig. 2. The numerical results show that the best fit
occurs in the range $r\in (1,9)$, hence we select this range to
perform multifractal analysis and get the spectrum of generalized
dimensions $D_{q}$.

After this spectrum has been obtained, we use $\Delta D(q)=\max
D(q)-\lim D(q)$ to verify how $D_{q}$ changes along each curve.
The quantity $\Delta D(q)$ has been used in the literature to
describe the density of an object. In this paper, based on our
modified fixed-size box covering method, $\Delta D(q)$ can help to
understand how the edge density changes in the complex network. In
other words, a larger value of $\Delta D(q)$ means the edge
distribution is more uneven. More specifically, for a network,
edge distribution could vary from an area of hubs where edges are
dense to an area where nodes are just connected with a few links.

In the following sections, we calculate the generalized fractal dimensions $%
D_{q}$. From the shape of $D_{q}$, we determine the
multifractality of the network using the method described above.
We then calculate $\Delta D(q)$ to verify how $D_{q}$ changes
along each curve.\textbf{\ }

\section{Results and discussions}

 In recent years, with the development of technology, the
research on networks has shifted away from the analysis of single
small graphs and the properties of individual vertices or edges
within such graphs to consideration of large-scale statistical
properties of complex networks. Newman$^{\cite{Newman2003}}$
reviewed some latest works on the structure and function of
networked systems such as the Internet, the World Wide Web, social
networks and a variety of biological networks. Besides reviewing
empirical studies, the author also focused on a number of
statistical properties of networks including path lengths, degree
distributions, clustering and resilience. In this paper, we pay
attention to another aspect\ of networks, namely their
multifractality. We aim to develop a tool based on this property
to characterize and classify real-world networks.

 It has been shown that many real complex networks share
distinctive characteristics that differ in many ways from random
and regular networks.$^{\cite{LeeJung2006, GuoCai2009,
Newman2003}}$ Fundamental properties of complex networks such
small-world effect and the scale-free degree distribution have
attracted much attention recently. These properties have in fact
been found in many naturally occurring networks. In Subsections
3.1, 3.2 and 3.3, we generate scale-free networks using the BA
model of Barabasi and Albert,$^{\cite{Barabsi1999}}$ small-world
networks using the NW model of Newman and
Watts,$^{\cite{Newman1999}}$ then random networks using the ER
model of Erd\"{o}s and R\'{e}nyi$^{\cite{ER1960}}$ respectively.
We then apply our modified fixed-size box counting algorithm to
analyze the multifractal behavior of these networks.

\subsection{Scale-free networks}

 We use the elegant and simple BA model of Barabasi and Albert$^{\cite%
{Barabsi1999}}$ to generate scale-free networks. The origin of the
scale-free behavior in many systems can be traced back to this BA
model, which correctly predicts the emergence of scaling exponent.
The BA model consists of two mechanisms : Initially, the network
begins with a seed network of $n$ nodes, where $n\geq 2$ and the
degree of each node in the initial network should be at least 1,
otherwise it will always remain disconnected from the rest of the
network. For example, here we start with an initial network of 5
nodes. Its interaction matrix is
$$
\left(
\begin{array}{ccccc}
0 & 1 & 0 & 0 & 1 \\
1 & 0 & 0 & 1 & 0 \\
0 & 0 & 0 & 1 & 0 \\
0 & 1 & 1 & 0 & 0 \\
1 & 0 & 0 & 0 & 0%
\end{array}%
\right) .
$$
We then add one node to this initial network at a time. Each new
node is connected to $n$ existing nodes with a probability that is
proportional to the number of links that the existing nodes
already have. Formally, the probability $p_{i}$ that the new node
is connected to node $i$ is
\begin{equation}
p_{i}=\frac{k_{i}}{\sum_{j}k_{j}},
\end{equation}%
where $k_{i}$ is the degree of node $i$. So hubs tend to quickly
accumulate even more links, while nodes with only a few links are
unlikely to be chosen as destination for a new link.

In this paper, these scale-free networks are generated based on
the same seed which is the initial network of 5 nodes. For better
comparison, in each step, one node will be added into the network
with one link. Then we apply the modified fixed-size box counting
method on them to detect their multifractal behavior.

In Fig. 3. we can see that scale-free networks are multifractal by
the shape of the $D_{q}$ curves. The $D_{q}$ functions of these
networks decrease sharply after the peak. An explanation is that,
in a scale-free network, there are several nodes which are known
as hubs that have a large number of edges connected to them, so
the edge density around the areas near the hubs is larger than the
remaining parts of the network.

We summarize the numerical results in Table 1 including the number
of nodes,
number of edges, diameter, power law exponent $\gamma $, maximum value of $%
D_{q}$, limit of $D_{q}$, and $\Delta D_{q}$. From these results
we could see that scale free networks with larger size (more nodes
and more edges) are likely to have larger values of the maximum
and limit of $D_{q}$. In other words, the function $D_{q}$
increases with the size of a scale-free network. An explanation
for this situation is that larger scale-free networks usually have
more hubs which make the structure of the network more complex.

Scale-free networks show a power-law degree distribution of
$P(k)\sim k^{-\gamma }$, where $P(k)$ is the probability of a node
randomly chosen with degree $k$. It was shown in Ref.
\cite{AJB1999, AB2002} that when $\gamma <2,$ the average degree
diverges; while for $\gamma
>3,$ the standard deviation of the degree diverges. It has been
found that the degree exponent $\gamma $ usually varies in the
range of $2<\gamma <3$ $^{\cite{AJB1999}}$ for most scale-free
networks. Accordingly, we computed the power-law exponent of these
generated scale-free networks. The results show that there doesn't
seem to be any
clear relationship between power law and the maximum of $D_{q},$ limit of $%
D_{q}$ or $\Delta D_{q}$.

\subsection{Small-world networks}

 In 1998, Watts and Strogatz$^{\cite{Watts1998}}$ proposed
a single-parameter small-world network model that bridges the gap
between a regular network and a random graph. With the WS
small-world model, one can link a regular lattice with pure random
network by a semirandom network with high clustering coefficient
and short average path length. Later on, Newman and
Watts$^{\cite{Newman1999}}$ modified the original WS model. In the
NW model, instead of rewiring links between nodes, extra links
called shortcuts are added between pairs of nodes chosen at
random, but no links are removed from the existing network. The NW
model is equivalent to the WS model for small $p$ and sufficiently
large $N$, but easier to proceed.

In this paper, we use the NW model as follows. Firstly, we should
select three parameters: the dimension $n$, which is the number of
nodes in a graph; the mean degree $k$ (assumed to be an even
integer), which is the number of nearest-neighbors to connect; and
the probability $p$ of adding a shortcut in a given row, where
$0\leq p\leq 1$ and $n\gg k\gg \ln (n)\gg 1$. Secondly, we follow
two steps:

\begin{enumerate}
\item[(i)] Construct a regular ring lattice. For example, if the
nodes are named $N_{0},...,N_{n-1}$, there is an edge $e_{ij}$
between node $N_{i}$ and $N_{j}$ if and only if $|i-j|\equiv K$
for $K\in \lbrack 0,k/2]$;

\item[(ii)] Add a new edge between nodes $N_{i}$ and $N_{j}$ with
probability $p$.
\end{enumerate}

An illustration of this generating process is given in Fig. 4. The
upper left figure corresponds to the probability $p=0$. It is a
regular network containing 20 nodes and each node has two
neighbors on both sides. In other words, in this regular network,
each node has four edges.  Then we start generating small-world
networks based on this regular network. The upper right figure of
Fig. 4 corresponds to the probability $p=0.1;$ one edge is added
into the network. The network then becomes a small-world network.
The bottom left figure corresponds to the probability $p=0.5;$
seven edges are added into the original regular network and it is
also a small-world network. The bottom right figure corresponds to
the probability $p=1$; 10 edges are added into the original
small-world network and this time it becomes a random network.

 In this paper, we firstly generated a regular network which
contains 5000 nodes and 250,000 edges. Each node has 50 edges on
each side. Then we apply the modified fixed-size box counting
method on this regular network. The numerical results are shown in
the last row of Table 2. Both the maximum value of $D_{q}$ and the
limit of $D_{q}$ are equal to one, thus $\Delta D_{q}=0$. This is
because regular networks are not fractal, and they have dimension
one. Secondly, for better comparison, we generated ten small-world
networks based on a regular network of 5000 nodes with 5 edges on
each side of a node. During the generation, when the probability
$p$ increases, more edges are added into the original regular
network. Then we apply the modified fixed-size box counting method
on them to detect their multifractal behavior. We summarize the
numerical results in Table 2, which includes the number of nodes,
number of edges, diameter, probability $p$ (the generating
parameter), maximum value of $D_{q}$ and $\Delta D_{q}$. These
results indicate that, when $p$ increases, more edges are added
and accordingly both the maximum and limit values of $D_{q}$
increase.

In Fig. 5 we can see that the $D_{q}$ curve of a regular network
whose
probability $p=0$ during generation is a straight line with the value of $1$%
. The $D_{q}$ curves of the other small-world networks are also
approximately straight lines but with different $D_{q}$ values. So
these networks are not multifractal. Another interesting property
is apparent when $0.03<p<0.2$, in which case $D_{q}$ increases
along with the value of $p$. More specifically, when $p$
increases, more edges are added to the network, and both the
maximum and limit values of $D_{q}$ and limit of $D_{q}$ increase.
The values of $\Delta D_{q}$ are all within the error range,
confirming that the $D_{q}$ curves are straight lines.

\subsection{Random networks}

 The Erd\"{o}s-R\'{e}nyi random graph
model$^{\cite{ER1960}}$ is the oldest and one of the most studied
techniques to generate complex networks.

We generate random networks based on the ER
model$^{\cite{ER1960}}$:

\begin{enumerate}
\item[(i)] Start with $N$ isolated nodes;

\item[(ii)] Pick up every pair of nodes and connect them by an
edge with probability $p.$
\end{enumerate}

Usually, the results of this generation are separated subnetworks.
In this work, we just consider the largest connected part as the
network to work on and apply the modified fixed-size box counting
method to detect their multifractal behaviors. We then summarize
the numerical results in Table 3 including the number of nodes,
number of edges, diameter, probability $p$
(the generating parameter), maximum value of $D_{q}$, limit of $D_{q}$, and $%
\Delta D_{q}$. These results indicate that there is no clear
relationship between $D_{q}$ and the size of the random network.

In Fig. 6, we can see that the $D_{q}$ curves of random networks
decrease slowly after the peak and the changes could be seen by
the values of $\Delta D_{q}$. This pattern occurs because, during
the generating process, nodes are randomly connected with
probability $p$, and few hubs may exist. Compared with scale-free
networks, this decrease supports the claim that, in random
networks, edges are distributed more symmetrically.\textbf{\ }

\textbf{\textbf{Remark}: }In the present study, we consider the
generalized fractal dimensions $D_{q}$ to determine whether the
object is multifactal from the shape of $D_{q}$. For a monofractal
system, which has the same
scaling behavior at any point, $D_{q}$ should be a constant independent of $%
q $, while for a multifractal, the $D_{q}$ should be a
non-increasing nonlinear curve as $q$ increases. However, in our
results, an anomalous
behavior is observed: the $D_{q}$ curves increase at the beginning when $q<0$%
. This anomalous behavior has also been observed in Bos {\it et
al.},$^{\cite{Bos1996}}$ Smith and Lange,$^{\cite{Smith1998}}$
Fern\'{a}ndez {\it et al.}.$^{\cite{Fern1999}}$ Some reasons for
this behavior have been suggested, including that the boxes
contain few elements,$^{\cite{Fern1999}}$ or the small scaling
regime covers less than a decade so that we cannot extrapolate the
box counting results for the partition function to zero box
size.$^{\cite{Bos1996}}$ In encountering the anomalous spectra of
$D_{q}$, we tried another method of multifractal analysis called
the sand-box method, but the linear regression fittings are not
satisfactory. We therefore used the modified fixed-size box
counting algorithm in this research. For the purpose of detecting
the multifractality of complex networks, we adopt the anomalous
spectra of $D_{q}$ and focus on the decreasing parts which are
presented in Figs. 3 to 7.

\subsection{Protein-protein interaction networks}

 Our fractal and multifractal analyses are based on
connected networks which do not have separated parts or isolated
nodes. In order to apply them to protein-protein interaction (PPI)
networks, some preparation is needed in advance. Firstly, we need
to find the largest connected part of each data set. For this
purpose many tools and methods could be used. In our study, we
adopt the Cytoscape$^{\cite{Cytoscape}}$ which is an open
bioinformatics software platform for visualizing molecular
interaction networks and analyzing network graphs of any kind
involving nodes and edges. In using Cytoscape, we could get the
largest connected part of each interacting PPI data set and this
connected part is the network on which fractal and multifractal
analyses are performed.

The protein-protein interaction data we used here are mainly
downloaded from two databases: The PPI networks of Drosophila
melanogaster (fruit fly), C. elegans, Arabidopsis thaliana and
Schizosaccharomyces pombe are downloaded from
BioGRID.$^{\cite{BioGRID}}$ The PPI networks of S. cerevisiae
(baker's yeast), E. coli and H. pylori are download from
DIP.$^{\cite{DIP}}$ We also use the same human PPI network data as
in Ref. \cite{LeeJung2009}.

We calculated the $D_{q}$ spectra for eight PPI networks of
different organisms as shown in Fig. 7. From these $D_{q}$ curves,
we see that all PPI networks are multifractal and there are two
clear groupings of organisms based on the peak values of their
$D_{q}$ curves. The first group includes human, Drosophila
melanogaster, S. cerevisiae, and C. elegans. The second group just
includes two bacteria E.coli and H. pylori. We also see that the
PPI networks of the eight organisms have similar shape for the
$D_{q}$ curves. They all increase when $q\in \lbrack 0,1]$, and
reach their peak values around $q=2$, then decrease sharply as
$q>2$ and finally reach their limit value when $q>10$. So we can
take $\lim D(q)=D(20)$ and use $\Delta D(q)=\max D(q)-\lim D(q)$
to verify how the $D_{q}$ function changes along each curve. We
summarize the corresponding numerical results in Table 4.

\section{Conclusions}

 After analyzing a variety of real complex networks, Song
{\it et al.}$^{\cite%
{Song2005}}$ found that they consist of self-repeating patterns on
all length scales, i.e., complex networks have self-similar
structures. They found that the box-counting method is a proper
tool to unfold the self-similar properties of complex networks and
to further investigate network properties.

However, describing objects by a single fractal dimension is a
limitation of fractal analysis, especially when the networks
exhibit a multifractal behavior. Multifractal analysis is a useful
way to characterize the spatial heterogeneity of both theoretical
and experimental fractal patterns. It allows the computation of a
set of fractal dimensions, especially the generalized fractal
dimensions $D_{q}$.

A modified algorithm for analyzing the multifractal behavior of
complex networks is introduced in this paper. This algorithm is
applied on generated scale-free networks, small-world networks and
random networks as well as protein-protein interaction networks.
The numerical results indicate that multifractality exists in
scale-free networks and PPI networks, while for small-world
networks and random networks their multifractality is not
clear-cut, particularly for small-world networks generated by the
NW model. Furthermore, for scale-free networks, the values of
$D_{q}$ increase when the size of the network increases because
larger scale-free networks usually have more hubs which make the
structure of the network more complex. However, for random
networks there is no clear relationship between $D_{q}$ and the
size of the network.\textbf{\ }The quantity $\Delta D(q)=\max
D(q)-\lim D(q)$ has been used to investigate how $D_{q}$ changes. Larger $%
\Delta D(q)$ means the network's edge distribution is more uneven;
while smaller $\Delta D(q)$ means the network's edge distribution
is more symmetrical, which is the case for random networks.

These results support that the algorithm proposed in this paper is
a suitable and effective tool to perform multifractal analysis of
complex
networks. Especially, in conjunction with the derived quantities from $D_{q}$%
, the method and algorithm provide a needed tool to cluster and
classify real networks such as the protein-protein interaction
networks of organisms.

\section*{Acknowledgements}

This project was supported by the Australian Research Council
(Grant No. DP0559807), the Natural Science Foundation of China
(Grant No. 11071282),  the Chinese Program for Changjiang Scholars
and Innovative Research Team in University (PCSIRT) (Grant No.
IRT1179), the Chinese Program for New Century Excellent Talents in
University (Grant No. NCET-08-06867), the Research Foundation of
Education Commission of Hunan Province of China (grant No.
11A122),  Hunan Provincial Natural Science Foundation of China
(Grant No. 10JJ7001), Science and Technology Planning Project of
Hunan province of China (Grant No. 2011FJ2011), the Lotus Scholars
Program of Hunan province of China, the Aid Program for Science
and Technology Innovative Research Team in Higher Education
Institutions of Hunan Province of China, and a China Scholarship
Council--Queensland University of Technology Joint Scholarship.

\newpage
\newpage

\begin{table}[tpb]
\caption{Comparison of different scale-free networks}
\begin{center}
{\footnotesize \
\begin{tabular}{c|c|c|c|c|c|c}
\hline Number of nodes & Number of edges & Diameter & $\gamma$ &
Max(Dq) & Lim(Dq) & $\Delta Dq$ \\ \hline
500 & 499 & 13 & 1.94 $\pm $ 0.02 & 2.67 & 1.36 & 1.31 \\
1000 & 999 & 16 & 2.02 $\pm $ 0.07 & 2.93 & 1.47 & 1.46 \\
1500 & 1499 & 17 & 2.09 $\pm $ 0.04 & 2.96 & 1.65 & 1.30 \\
2000 & 1999 & 20 & 1.99 $\pm $ 0.08 & 3.05 & 1.76 & 1.29 \\
3000 & 2999 & 20 & 2.06 $\pm $ 0.04 & 3.26 & 1.83 & 1.44 \\
4000 & 3999 & 23 & 2.09 $\pm $ 0.03 & 3.32 & 1.80 & 1.52 \\
5000 & 4999 & 23 & 2.08 $\pm $ 0.04 & 3.26 & 1.75 & 1.51 \\
6000 & 5999 & 22 & 2.06 $\pm $ 0.04 & 3.39 & 1.88 & 1.51 \\
7000 & 5999 & 28 & 2.08 $\pm $ 0.04 & 3.39 & 2.10 & 1.29 \\
8000 & 5999 & 25 & 1.91 $\pm $ 0.12 & 3.33 & 2.11 & 1.22 \\ \hline
\end{tabular}
}
\end{center}
\end{table}

\begin{table}[tpb]
\caption{Comparison of different small-world networks and regular
networks with 5000 nodes}
\begin{center}
{\footnotesize \
\begin{tabular}{c|c|c|c|c|c|c}
\hline
Number of nodes & Number of edges & Diameter & p & Max(Dq) & Lim(Dq) & $%
\Delta Dq$ \\ \hline
5000 & 25159 & 33 & 0.03 & 2.31 & 2.28 & 0.03 \\
5000 & 25207 & 29 & 0.04 & 2.43 & 2.37 & 0.06 \\
5000 & 25290 & 23 & 0.06 & 2.56 & 2.53 & 0.03 \\
5000 & 25358 & 23 & 0.08 & 2.66 & 2.63 & 0.03 \\
5000 & 25513 & 18 & 0.1 & 2.81 & 2.75 & 0.06 \\
5000 & 25621 & 15 & 0.13 & 2.89 & 2.83 & 0.06 \\
5000 & 25792 & 15 & 0.15 & 2.99 & 2.93 & 0.06 \\
5000 & 26017 & 12 & 0.2 & 3.08 & 3.04 & 0.04 \\
regular network 5000 & 250000 & 50 & 0 & 1 & 1 & 0.00 \\ \hline
\end{tabular}
}
\end{center}
\end{table}

\begin{table}[tpb]
\caption{Comparison of different random networks}
\begin{center}
{\footnotesize \
\begin{tabular}{c|c|c|c|c|c|c}
\hline
Number of nodes & Number of edges & Diameter & p & Max(Dq) & Lim(Dq) & $%
\Delta Dq$ \\ \hline
449 & 610 & 15 & 0.005 & 2.42 & 2.14 & 0.28 \\
994 & 2502 & 8 & 0.005 & 3.32 & 2.87 & 0.45 \\
1991 & 5939 & 9 & 0.003 & 3.73 & 3.41 & 0.32 \\
2484 & 6310 & 11 & 0.002 & 3.70 & 3.33 & 0.37 \\
2790 & 4374 & 18 & 0.001 & 3.29 & 2.95 & 0.34 \\
3373 & 5978 & 15 & 0.001 & 3.47 & 3.15 & 0.32 \\
3931 & 8125 & 13 & 0.001 & 3.67 & 3.35 & 0.31 \\
4919 & 10179 & 13 & 0.0008 & 3.78 & 3.39 & 0.39 \\
5620 & 8804 & 16 & 0.00058 & 3.54 & 3.21 & 0.33 \\ \hline
\end{tabular}
}
\end{center}
\end{table}

\begin{table}[tpb]
\caption{Comparison of different PPI networks}
\begin{center}
{\footnotesize \
\begin{tabular}{c|c|c|c|c|c|c|c}
\hline Networks & Number of nodes & Number of edges & Diameter &
$D_0$ & Max(Dq) & Lim(Dq) & $\Delta Dq$ \\ \hline
Human & 8934 & 41341 & 14 & 2.34 & 4.89 & 2.65 & 2.24 \\
Drosophila Melanogaster & 7476 & 26534 & 11 & 2.34 & 4.84 & 2.87 & 1.97 \\
S. cerevisiae & 4976 & 21875 & 10 & 2.36 & 4.62 & 2.48 & 2.14 \\
E.coli & 2516 & 11465 & 12 & 2.14 & 4.15 & 2.10 & 2.05 \\
H.pylori & 686 & 1351 & 9 & 2.27 & 3.47 & 1.91 & 1.56 \\
C.elegans & 3343 & 6437 & 13 & 2.28 & 4.47 & 1.49 & 2.98 \\
Arabidopsis Thaliana & 1298 & 2767 & 25 & 1.83 & 2.51 & 1.62 &
0.89 \\ \hline
\end{tabular}
}
\end{center}
\end{table}

\begin{figure}[tbp]
\centerline{\includegraphics[width=0.6\textwidth]{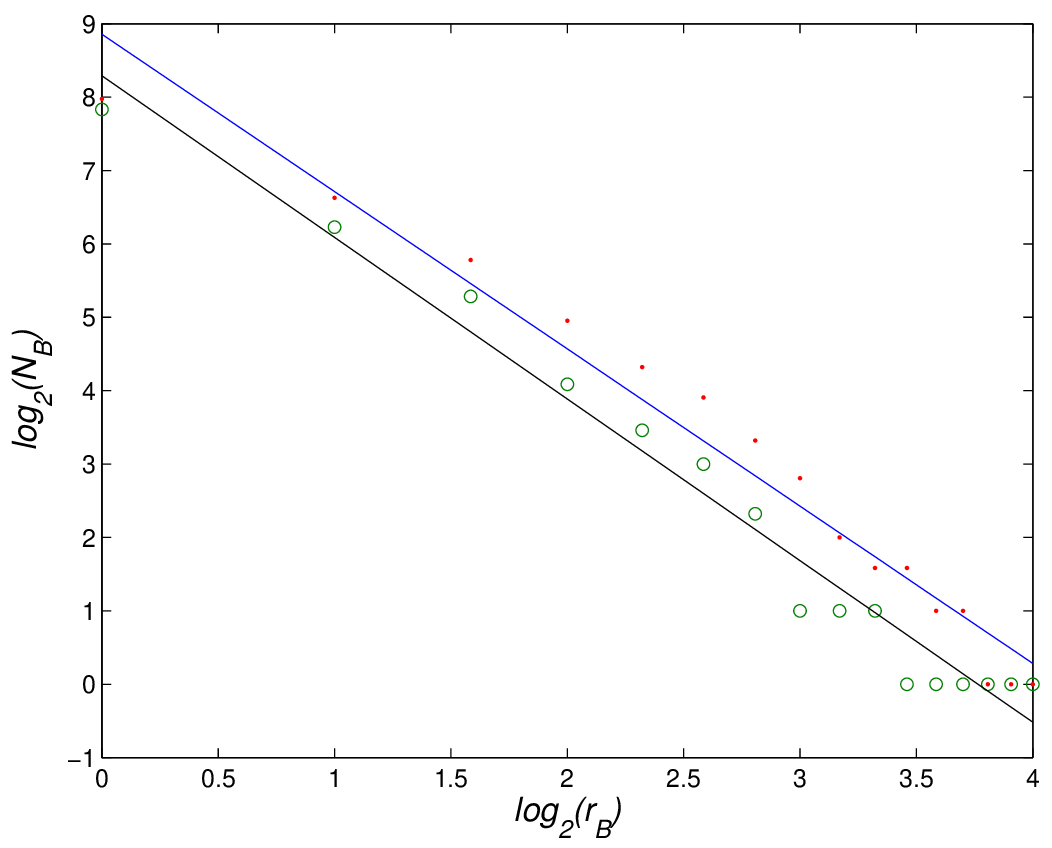}}
\caption{Fractal scaling of the human PPI network (o) and its
skeleton (.). The fractal dimension is the absolute value of the
slope of each linear fit, which is 2.20 $\pm $ 0.09 for the
original network and 2.07 $\pm $ 0.09 for its skeleton.}
\end{figure}

\begin{figure}[tbp]
\centerline{\includegraphics[width=0.6\textwidth]{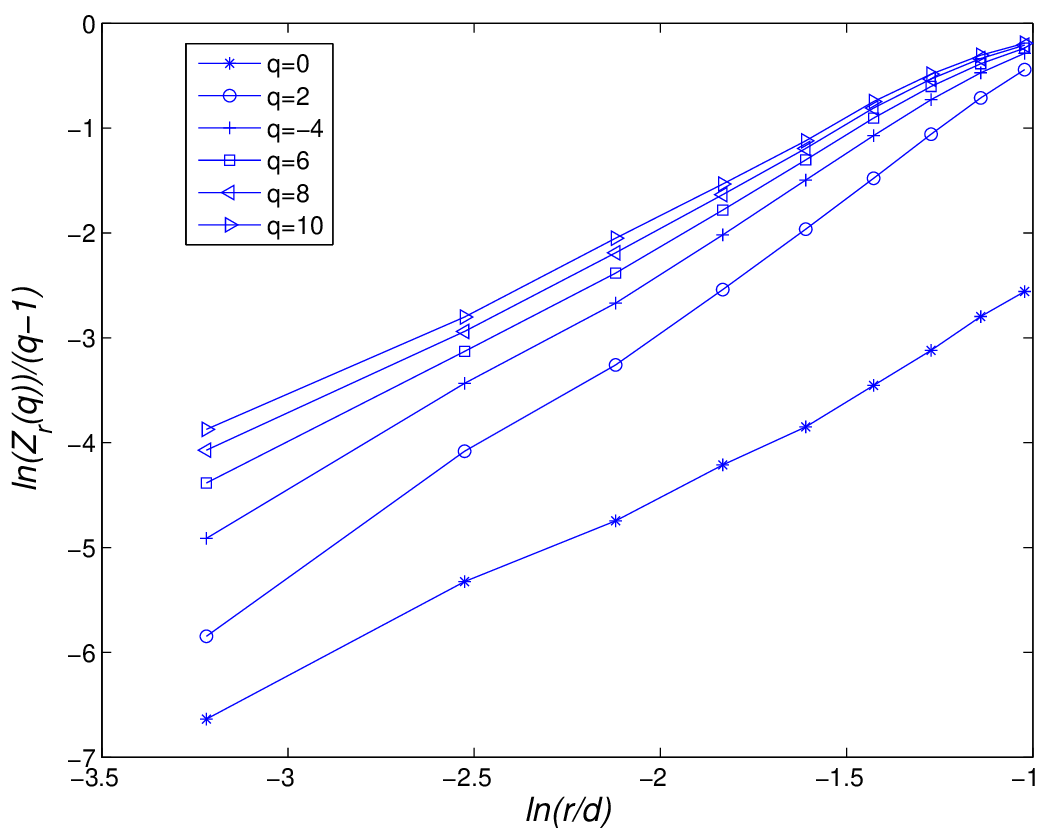}}
\caption{Linear regressions of the Arabidopsis Thaliana PPI
network.}
\end{figure}

\begin{figure}[tbp]
\centerline{\includegraphics[width=0.6\textwidth]{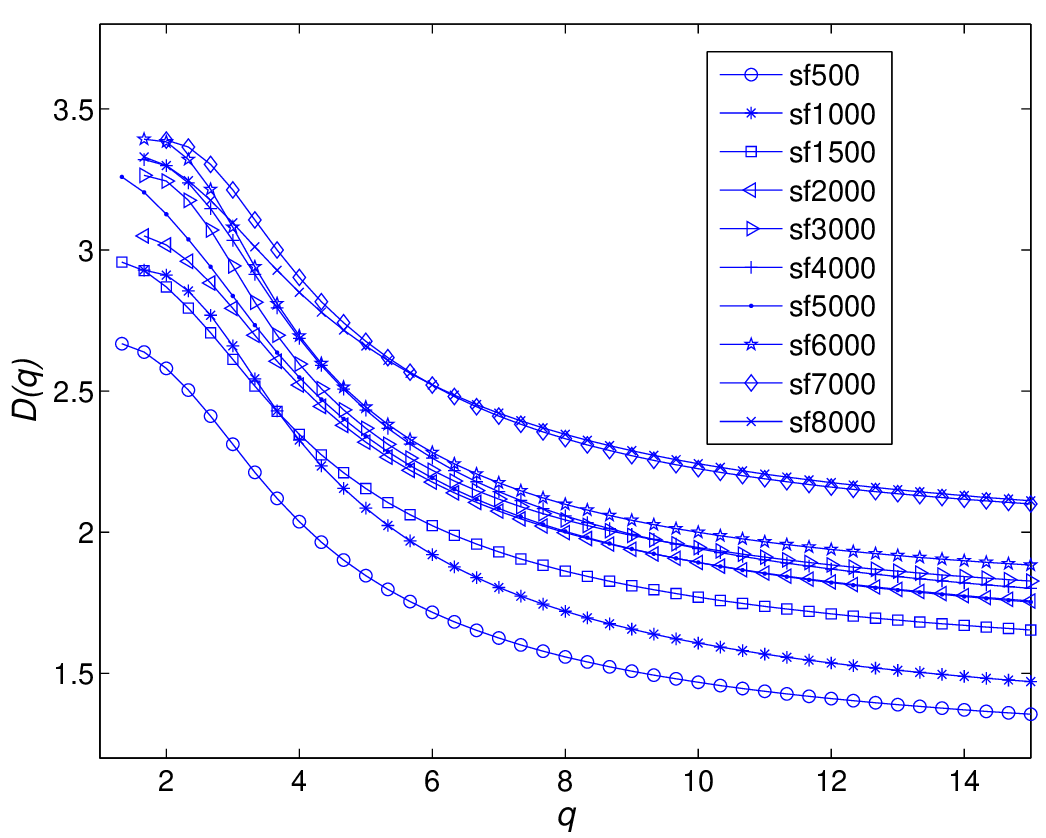}}
\caption{The $D_{q}$ curves for theoretically generated scale-free
networks.}
\end{figure}

\begin{figure}[tbp]
\centerline{\includegraphics[width=0.6\textwidth]{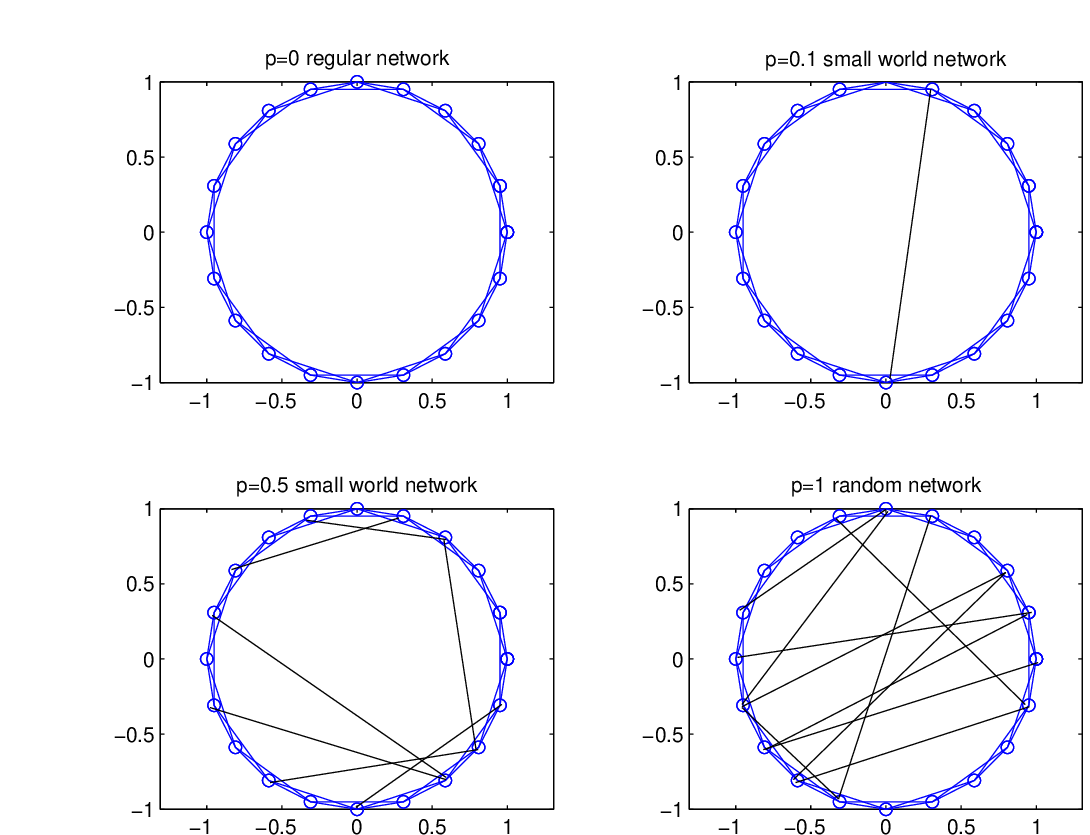}}
\caption{Generation of small-world networks \ by the NW model.}
\end{figure}

\begin{figure}[tbp]
\centerline{\includegraphics[width=0.6\textwidth]{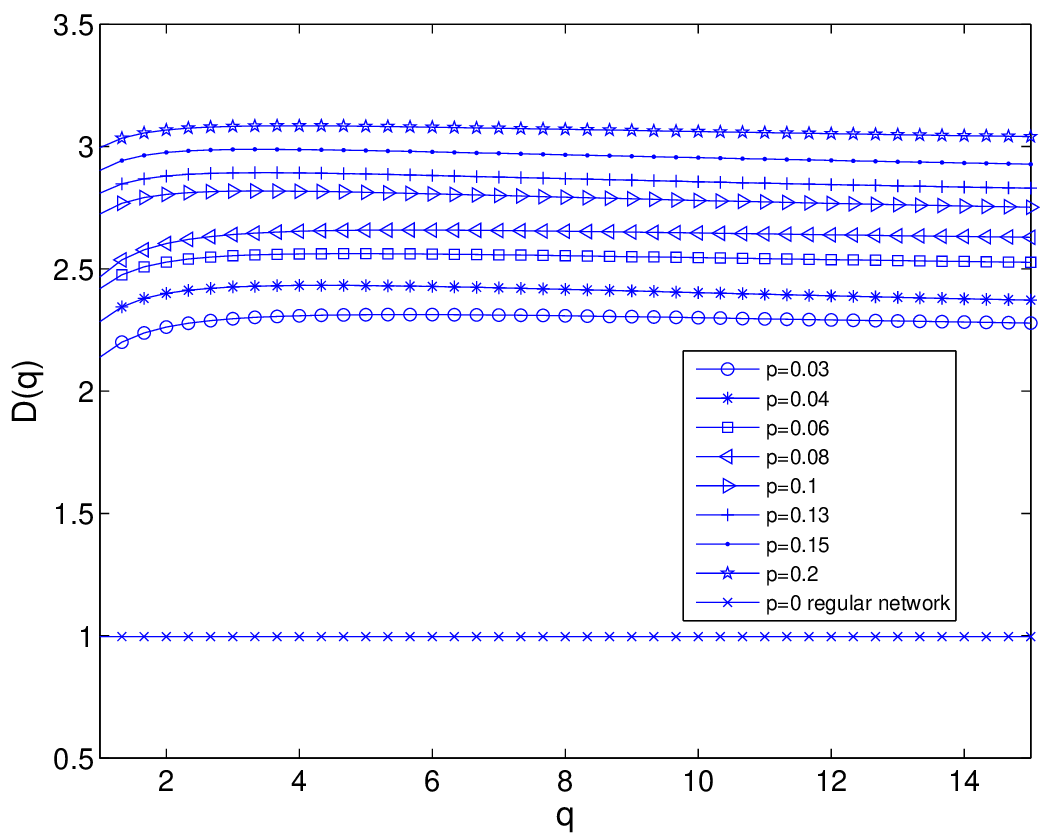}}
\caption{The $D_{q}$ curves for theoretically generated
small-world networks. }
\end{figure}

\begin{figure}[tbp]
\centerline{\includegraphics[width=0.6\textwidth]{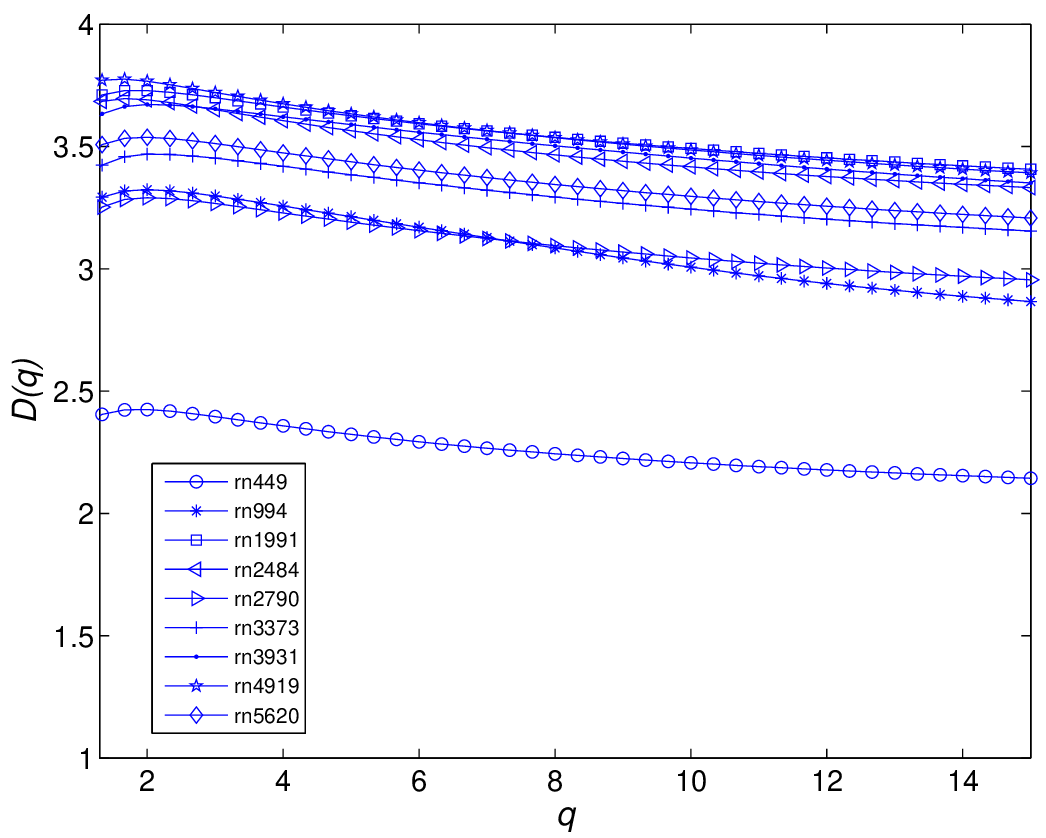}}
\caption{The $Dq$ curves for different random networks.}
\end{figure}

\begin{figure}[tbp]
\centerline{\includegraphics[width=0.6\textwidth]{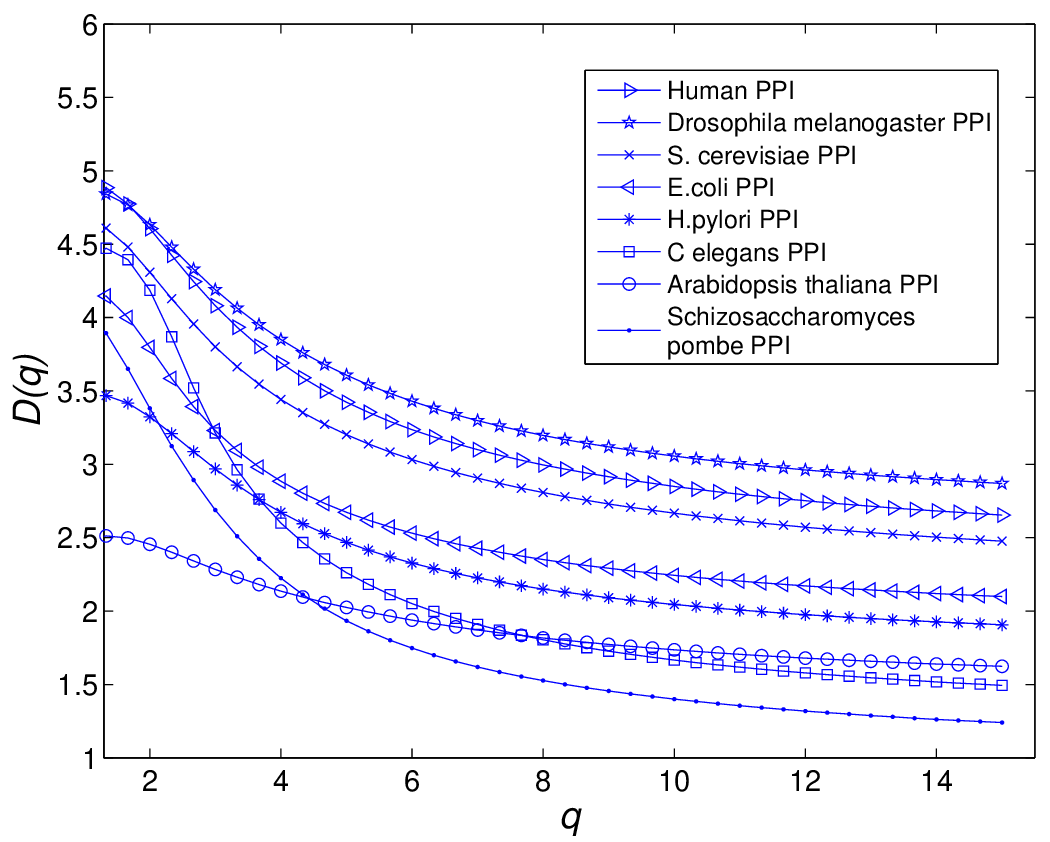}}
\caption{The $D_{q}$ curves for different PPI networks.}
\end{figure}


\begin{thebibliography}{10}

{\footnotesize

\bibitem[1]{Song2005} Song C, Havlin S and  Makse H A, 2005 \emph{Nature} \textbf{433} 392

\bibitem[2]{LeeJung2006}  Lee C Y and Jung S  2006 \emph{Phys. Rev. E} \textbf{73}
 066102

\bibitem[3]{GuoCai2009}  Guo L and  Cai X 2009 \emph{Chin. Phys. Lett.} \textbf{26}
 088901

\bibitem[4]{ER1960}  Erdos P and Renyi A 1960 \emph{Publ. Math. Inst. Hung. Acad. Sci.}
\textbf{5} 17

\bibitem[5]{Milgram1967}  Milgram S 1967 \emph{Psychol.
Today} \textbf{2} 60

\bibitem[6]{AJB1999} Albert R, Jeong H and  Barabasi A L 1999 \emph{Nature}
\textbf{401} 130

\bibitem[7]{AB2002}  Albert R and  Barabasi A L 2002 \emph{Rev. Mod. Phys.}
 \textbf{74} 47

\bibitem[8]{Faloutsos1999}  Faloutsos M,  Faloutsos P and Faloutsos C 1999
\emph{Comput. Commun. Rev.} \textbf{29}  251

\bibitem[9]{Mandelbrot1983} Mandelbrot B B 1983 \emph{The fractal Geometry of
Nature} (New York, Academic Press)

\bibitem[10]{Feder1988}  Feder J 1988 \emph{Fractals} (New York, Plenum)

\bibitem[11]{Falconer1997}  Falconer K 1997 \emph{Techniques in Fractal Geometry}
(New York, Wiley)

\bibitem[12]{Eguiluz2003} Eguiluz V M,  Hernandez-Garcia E,
Piro O and  Klemm K 2003 \emph{Phys. Rev. E} \textbf{68} 055102(R)

\bibitem[13]{Song2007}  Song C,  Gallos L K,
Havlin S and  Makse H A 2007 \emph{J. Stat. Mech.: Theor. Exper.}
\textbf{3} P03006

\bibitem[14]{Kim2007}  Kim J S, Goh K I, Salvi G, Oh E,  Kahng B, and
Kim D 2007 \emph{Phys. Rev. E} \textbf{75} 016110

\bibitem[15]{ZhouJiang2007}  Zhou W X,  Jiang Z Q and Sornette D
2007 \emph{Physica A} \textbf{375} 741

\bibitem[16]{GaoHu2008} Gao L,  Hu Y and  Di Z 2008 \emph{Phys. Rev. E} \textbf{78}
 046109

\bibitem[17]{Shanker2007}  Shanker O 2007 \emph{Mod. Phys. Lett.B} \textbf{21} 321

\bibitem[18]{Rozenfeld2009}  Rozenfeld H D and  Makse H A 2009
\emph{Chem. Engneer. Sci.} \textbf{64} 4572

\bibitem[19]{Rozenfeld2010} Rozenfeld H D, Song C and
Makse H A 2010 \emph{Phys. Rev. Lett.} \textbf{104} 025701

\bibitem[20]{LiuKong2010} Liu J X and Kong X M 2010
\emph{Acta Physica Sinica} \textbf{59(4)} 2244

\bibitem[21]{Grassberger1983}  Grassberger P and Procaccia I 1983
\emph{Phys. Rev. Lett.} \textbf{50} 346

\bibitem[22]{Halsy1986}  Halsey T C,  Jensen M H,  Kadanoff L P,
Procaccia I, and  Shraiman B I 1986 \emph{Phys. Rev. A}
\textbf{33} 1141

\bibitem[23]{can00} Canessa E 2000 \emph{J.
Phys. A: Math. Gen.} \textbf{33} 3637

\bibitem[24]{Anh2000}  Anh V V,  Tieng Q M and  Tse Y K 2000 \emph{%
Int. Trans. Oper. Res.} \textbf{7} 349

\bibitem[25]{YuAnhLau2001a} Yu Z G,  Anh V and  Lau K S 2001
 \emph{Physica A} \textbf{301} 351

\bibitem[26]{YuAnhLau2001b} Yu Z G,  Anh V and  Lau K S  2001
\emph{Phys. Rev. E} \textbf{64} 31903

\bibitem[27]{YuAnhLau2001c} Yu Z G,  Anh V  and  Wang B 2001
\emph{Phys. Rev. E} \textbf{63} 11903

\bibitem[28]{AnhLauYu2002}  Anh V,  Lau K S and Yu Z G 2002
\emph{Phys. Rev. E} \textbf{66} 031910

\bibitem[29]{YuAnhLau2003} Yu Z G,  Anh V and  Lau K S 2003
\emph{Phys. Rev. E} \textbf{68} 021913

\bibitem[30]{YuAnhLau2004} Yu Z G,  Anh V and  Lau K S 2004 \emph{J. Theor.
Biol.} \textbf{226} 341

\bibitem[31]{ZhouYuDeng2005}  Zhou L Q, Yu Z G,  Deng J Q,  Anh V and
Long S C 2005 \emph{J. Theor. Biol.} \textbf{232} 559

\bibitem[32]{YuAnhLau2006} Yu Z G,  Anh V,  Lau K S  and  Zhou L Q 2006
\emph{Phys. Rev. E} \textbf{73} 031920

\bibitem[33]{YuXiao2010}  Yu Z G, Xiao Q J, Shi L,
Yu J W, and  Anh V 2010 {\em Chin. Phys. B} {\bf 19 (6)} 068701

\bibitem[34]{ZhuYu2011} Zhu S M, Yu Z G,  Anh V 2011
  {\em Chin. Phys. B} {\bf 20 (1)} 010505

\bibitem[35]{HanFu2010} Han J J and  Fu W J 2010 {\em Chin.
Phys. B} {\bf 19 (1)} 010205

\bibitem[36]{Kantelhardt2006}  Kantelhardt J W, Koscielny-Bunde E,
Rybski D,  Braun P, Bunde A and  Havlin S 2006 \emph{J. Geophys.
Res.} \textbf{111} D01106

\bibitem[37]{Veneziano2006}  Veneziano D,  Langousis A and  Furcolo P 2006
\emph{Water Resour. Res.} \textbf{42} W06D15

\bibitem[38]{Venugopal2006}  Venugopal V,  Roux S G,  Foufoula-Georgiou E and
 Arneodo A 2006 \emph{Water Resour. Res.} \textbf{42} W06D14

\bibitem[39]{YAWW2007} Yu Z G,  Anh V,  Wanliss J A and  Watson S M 2007
 \emph{Chaos, Solitons and Fractals} \textbf{31} 736

\bibitem[40]{ZangShang2007}  Zang B J and  Shang P J 2007
{\em Chin. Phys.} {\bf 16 (3)} 565

\bibitem[41]{YuAnhEastes2009} Yu Z G,  Anh V and  Eastes R 2009
 \emph{J. Geophys. Res.} \textbf{114} A05214

\bibitem[42]{YuAnhWang2010} Yu Z G,  Anh V,   Wang Y, Mao D and  Wanliss J 2010
\emph{J. Geophys. Res.} \textbf{115} A10219

\bibitem[43]{Polla2010}  Polla G,  Lovasz L and  Vicsek T 2010
\emph{Proc. Natl. Acad. Sci. U.S.A.} \textbf{107} 7640

\bibitem[44]{Dijkstra1959}
Dijkstra E W 1959 \emph{Numerische Mathematik} \textbf{1} 269

\bibitem[45]{Newman2003}  Newman M E J 2003 \emph{SIAM Rev.} \textbf{45} 167

\bibitem[46]{Barabsi1999}  Barabsi A L and  Albert R 1999 \emph{Science}
 \textbf{286} 509

\bibitem[47]{Newman1999} Newman M E J and  Watts D J 1999 \emph{Phys. Lett. A}
\textbf{263} 341

\bibitem[48]{Watts1998}  Watts D J and  Strogatz S H 1998
 \emph{Nature} \textbf{393} 440

\bibitem[49]{Bos1996}  Opheusden J H H,  Bos M T A, and
van der Kaaden G 1996 \emph{Physica A} \textbf{227} 183

\bibitem[50]{Smith1998} Smith T G and  Lange G D 1998 in \emph{%
Fractal in Biology and Medicine} (Eds: Nonnenmacher T F, Losa G A
and Weibel E R) (Basel, Birkh\"{a}user)

\bibitem[51]{Fern1999}  Fern$\acute{a}$ndez E,  Bolea J A,  Ortega G and
Louis E 1999 \emph{\ J. Neurosci. Methods} \textbf{89} 151

\bibitem[52]{Cytoscape} Cytoscape software:
http://cytoscapeweb.cytoscape.org/

\bibitem[53]{BioGRID} BioGRID: http://thebiogrid.org/download.php

\bibitem[54]{DIP} DIP: http://dip.doe-mbi.ucla.edu/dip/Download.cgi

\bibitem[55]{LeeJung2009}  Lee E,  Jung H,  Radivojac P,  Kim J W and
Lee D 2009 \emph{BMC Bioinformatics} \textbf{10} (Suppl 2) S2

}
\end{thebibliography}
\end{document}